\documentclass[12pt]{article}

\newcommand{\text}{\rm}

\begin{document}

\title{\textbf{The anomalous dimension of the composite operator }$A^{2}$ \textbf{%
in the Landau gauge}}
\author{D. Dudal\thanks{%
Research Assistant of The Fund For Scientific Research-Flanders, Belgium.},\
\ H. Verschelde \thanks{%
david.dudal@rug.ac.be, henri.verschelde@rug.ac.be} \\
{\small {\textit{Ghent University }}}\\
{\small {\textit{Department of Mathematical Physics and Astronomy,
Krijgslaan 281-S9, }}}\\
{\small {\textit{B-9000 GENT, BELGIUM}}} \and and S.P. Sorella\thanks{%
sorella@uerj.br} \\
{\small {\textit{UERJ - Universidade do Estado do Rio de Janeiro,}}} \\
{\small {\textit{\ Rua S\~{a}o Francisco Xavier 524, 20550-013 Maracan\~{a}, 
}}} {\small {\textit{Rio de Janeiro, Brazil.}}}}
\maketitle

\begin{abstract}
The local composite operator $A^{2}\;$is analysed in pure Yang-Mills theory
in the Landau gauge within the algebraic renormalization. It is proven that
the anomalous dimension of $A^{2}$ is not an independent parameter, being
expressed as a linear combination of the gauge $\beta $ function and of the
anomalous dimension of the gauge fields.
\end{abstract}

\vfill\newpage \ \makeatother

\renewcommand{\theequation}{\thesection.\arabic{equation}}

\section{Introduction}

Nowadays an increasing evidence has been reported on the relevance of the
local composite operator $A_{\mu }^{a}A^{a\mu \;}$for the nonperturbative
regime of Yang-Mills theories quantized in the Landau gauge. That this
operator has a special meaning in the Landau gauge can be simply understood
by observing that, due to the transversality condition $\partial _{\mu
}A^{a\mu }=0$, the integrated mass dimension two operator $\left( VT\right)
^{-1}\int d^{4}xA_{\mu }^{a}A^{\mu a}$ is gauge invariant, $VT$ being the
space-time volume. Lattice simulations \cite{b} have indeed provided strong
indications for the existence of the pure gluon condensate $\left\langle
A_{\mu }^{a}A^{a\mu \;}\right\rangle $, confirming its relevance for the
infrared dynamics of Yang-Mills. Also, the monopole condensation in compact $%
QED$ turns out to be related to a phase transition for this condensate \cite
{g}.

Recently, a renormalizable effective potential for $\left\langle A_{\mu
}^{a}A^{a\mu \;}\right\rangle $ has been obtained in \cite{v} by using the
local composite operator (LCO) technique \cite{v1}. This result shows that
the vacuum of pure Yang-Mills theory favors a nonvanishing value of this
condensate, which provides effective masses for the gluons while
contributing to the dimension four condensate $\left\langle \alpha
F^{2}\right\rangle $ through the trace anomaly. It is worth remarking here
that this mass has a pure dynamical origin and manifests itself without
breaking the gauge group. Both features are indeed expected in a confining
gauge theory, being in agreement with the Kugo-Ojima criterion for color
confinement \cite{ko}.

An important ingredient in the analysis of the effective potential for the
gluon condensate is the anomalous dimension $\gamma _{A^{2}}$ of the
operator $A_{\mu }^{a}A^{a\mu \;}$. Till now, $\gamma _{A^{2}}$ has been
computed up to three loops in the $\overline{MS}$ renormalization scheme 
\cite{gr}. The available expression for $\gamma _{A^{2}}$ shows rather
interesting properties concerning the operator $A_{\mu }^{a}A^{a\mu \;}$in
the Landau gauge. It turns out indeed that, besides being multiplicative
renormalizable, its anomalous dimension can be expressed as a combination of
the gauge $\beta $ function and of the anomalous dimension $\gamma _{A}$ of
the gauge fields, according to the relation 
\begin{equation}
\gamma _{A^{2}}=-\left( \frac{\beta (a)}{a}+\gamma _{A}(a)\right)
\;,\;\;\;\;\;\;\;a=\frac{g^{2}}{16\pi ^{2}}\;,  \label{ri}
\end{equation}
which can be easily checked up to three loops \cite{gr}. This feature
strongly supports the existence of some underlying Ward identities which
should be at the origin of eq.$\left( \ref{ri}\right) $, meaning that $%
\gamma _{A^{2}}$ is not an independent parameter of the theory.

The aim of this paper is to provide an affirmative answer concerning the
possibility of giving a purely algebraic proof of the relation $\left( \ref
{ri}\right) $, which extends to all orders of perturbation theory. Our proof
will rely only on the use of the Slavnov-Taylor identity and of an
additional Ward identity, known as the ghost Ward identity, present in the
Landau gauge \cite{bps}. Furthermore, according to \cite{bps}, it turns out
that also the anomalous dimension $\gamma _{c}$ of the Faddeev-Popov ghosts
can be written as a combination of $\beta $ and $\gamma _{A}$, namely

\begin{equation}
2\gamma _{c}(a)=\frac{\beta (a)}{a}-\gamma _{A}(a)\;.  \label{ri1}
\end{equation}
Both relations $\left( \ref{ri}\right) $ and $\left( \ref{ri1}\right) $ can
be used as a very useful check for higher order computations in Yang-Mills
theories quantized in gauges which reduce to the Landau gauge when the gauge
parameters are set to zero, as in the case of the nonlinear Curci-Ferrari
gauge \cite{gr}.

The work is organized as follows. In Sect.1 we give a brief account of eqs.$%
\left( \ref{ri}\right) $ and $\left( \ref{ri1}\right) $ by making use of the
available three loops expressions. Sect.2 is devoted to their algebraic
proof.

\section{The anomalous dimension of the operator $A^{2}\;$in the Landau gauge
}

In order to give a short account of the relations $\left( \ref{ri}\right) $
and $\left( \ref{ri1}\right) $, let us recall the three loop expressions of
the gauge $\beta $ function and of the gauge and ghost fields anomalous
dimensions $\gamma _{A}$ and $\gamma _{c}$, as given in \cite{gr}. They read

\begin{equation}
\frac{\beta (a)}{a}=-\frac{11}{3}\left( Na\right) -\frac{34}{3}\left(
Na\right) ^{2}-\frac{2857}{54}\left( Na\right) ^{3}\;,  \label{beta}
\end{equation}
\begin{equation}
\gamma _{A}=-\frac{13}{6}\left( Na\right) -\frac{59}{8}\left( Na\right) ^{2}+%
\frac{\left( 648\varsigma (3)-39860\right) }{1152}\left( Na\right) ^{3}\;,
\label{ga}
\end{equation}
and 
\begin{equation}
\gamma _{c}=-\frac{3}{4}\left( Na\right) -\frac{95}{48}\left( Na\right) ^{2}-%
\frac{\left( 1944\varsigma (3)+63268\right) }{6912}\left( Na\right) ^{3}\;,
\label{gc}
\end{equation}
where $N$ is the number of colors corresponding to the gauge group $SU(N)$.
Making use of the relation 
\begin{equation}
\gamma _{A^{2}}=-\left( \frac{\beta (a)}{a}+\gamma _{A}(a)\right) \;,
\label{ga2}
\end{equation}
for the anomalous dimension of $A_{\mu }^{a}A^{a\mu \;}$one gets, up to the
third order,

\begin{itemize}
\item  first order 
\begin{equation}
\gamma _{A^{2}}^{(1)}=-\left( \frac{\beta ^{(1)}}{a}+\gamma
_{A}^{(1)}\right) =\frac{35}{6}\left( Na\right) \;.  \label{aa1}
\end{equation}

\item  second order 
\begin{equation}
\gamma _{A^{2}}^{(2)}=-\left( \frac{\beta ^{(2)}}{a}+\gamma
_{A}^{(2)}\right) =\frac{449}{24}\left( Na\right) ^{2}\;.  \label{aa2}
\end{equation}

\item  third order
\end{itemize}

\begin{equation}
\gamma _{A^{2}}^{(3)}\;=-\left( \frac{\beta ^{(3)}}{a}+\gamma
_{A}^{(3)}\right) =\left( \frac{75607}{864}-\frac{9}{16}\varsigma (3)\right)
\left( Na\right) ^{3}\;.  \label{a23}
\end{equation}
Expressions $\left( \ref{aa1}\right) $, $\left( \ref{aa2}\right) $, $\left( 
\ref{a23}\right) $ are in complete agreement with those given in\footnote{%
Notice that the anomalous dimension $\gamma _{O}$ for $A^{2}$ given in \cite
{gr} is related to $\gamma _{A^{2}}$ in eq.$\left( \ref{ga2}\right) $ by $%
\gamma _{A^{2}}=-4\gamma _{O}$.} \cite{gr}. Concerning now the ghost
anomalous dimension $\gamma _{c}$ in eq.$\left( \ref{gc}\right) $, it is
straightforward to verify in fact that the relation 
\begin{equation}
2\gamma _{c}(a)=\frac{\beta (a)}{a}-\gamma _{A}(a)\;,  \label{rel}
\end{equation}
holds. This equation expresses the nonrenormalization properties of the
ghost fields in the Landau gauge and, as shown in \cite{bps}, follows from
the ghost Ward identity. Although we are considering only pure Yang-Mills
theory, it is worth mentioning that eqs.$\left( \ref{ga2}\right) $ and $%
\left( \ref{rel}\right) $ remain valid also in the presence of matter
fields, as one can verify from \cite{gr}.

\section{Algebraic proof}

In this Section we provide an algebraic proof of the relation $\left( \ref
{ga2}\right) $. We shall make use of a suitable set of Ward identities which
can be derived in the Landau gauge in order to characterize the local
operator $A^{2}$. Let us begin by reminding the expression of the pure
Yang-Mills action in the Landau gauge

\begin{eqnarray}
S &=&S_{YM}+S_{GF+FP}  \label{lg} \\
&=&-\frac{1}{4}\int d^{4}xF_{\mu \nu }^{a}F^{a\mu \nu }+\int d^{4}x\left(
b^{a}\partial _{\mu }A^{a\mu }+\overline{c}^{a}\partial ^{\mu }D_{\mu
}^{ab}c^{b}\right) \;,  \nonumber
\end{eqnarray}
where 
\begin{equation}
D_{\mu }^{ab}\equiv \partial _{\mu }\delta ^{ab}+gf^{acb}A_{\mu }^{c}\;.
\label{dudal0}
\end{equation}
To study the operator $A_{\mu }^{a}A^{a\mu }$, we introduce it in the action
by means of a set of external sources. It turns out that three external
sources $J,$ $\eta ^{\mu }$ and $\tau ^{\mu }$ are needed, so that

\begin{eqnarray}
S_{\mathrm{J}} &=&\int d^{4}x\left[ J\frac{1}{2}A_{\mu }^{a}A^{a\mu }+\frac{%
\xi }{2}J^{2}-\eta ^{\mu }A_{\mu }^{a}c^{a}-\tau ^{\mu }s(A_{\mu
}^{a}c^{a})\right] \;  \label{lco} \\
&=&\int d^{4}x\left[ J\frac{1}{2}A_{\mu }^{a}A^{a\mu }+\frac{\xi }{2}%
J^{2}-\eta ^{\mu }A_{\mu }^{a}c^{a}+\tau ^{\mu }\partial _{\mu }c^{a}c^{a}+%
\frac{g}{2}\tau ^{\mu }f^{abc}A_{\mu }^{a}c^{b}c^{c}\right] \;,  \nonumber
\end{eqnarray}
where $s$ denotes the BRST\ operator acting as 
\begin{eqnarray}
sA_{\mu }^{a} &=&-D_{\mu }^{ab}c^{b}  \nonumber \\
sc^{a} &=&\frac{g}{2}f^{abc}c^{b}c^{c}  \nonumber \\
s\overline{c}^{a} &=&b^{a}  \nonumber \\
sb^{a} &=&-Jc^{a}\;  \nonumber \\
sJ &=&0\;  \nonumber \\
s\eta ^{\mu } &=&\partial ^{\mu }J\;  \nonumber \\
s\tau ^{\mu } &=&\eta ^{\mu }\;.  \label{s}
\end{eqnarray}
It is easy to check that

\begin{equation}
s\left( S_{YM}+S_{GF+FP}+S_{\mathrm{J}}\right) =0\;.  \label{sinv}
\end{equation}
According to the LCO procedure \cite{v,v1}, the dimensionless parameter $\xi 
$ is needed to account for the divergences present in the vacuum Green
function $\left\langle A^{2}(x)A^{2}(y)\right\rangle ,$ which are
proportional to $J^{2}$.

\subsection{Ward identities}

In order to translate the BRST\ invariance $\left( \ref{sinv}\right) $ into
the corresponding Slavnov-Taylor identity \cite{book}, we introduce two
additional external sources $\Omega _{\mu }^{a}\;$and $L^{a}\;$coupled to
the nonlinear BRST\ variation of $A_{\mu }^{a}\;$and $c^{a}$

\begin{equation}
S_{ext}=\int d^{4}x\left[ -\Omega ^{a\mu }D_{\mu }^{ab}c^{b}\;+L^{a}\frac{g}{%
2}f^{abc}c^{b}c^{c}\right] \;,  \label{om}
\end{equation}
with

\[
s\Omega _{\mu }^{a}=sL^{a}=0\;. 
\]
The complete action 
\begin{equation}
\Sigma =S_{YM}+S_{GF+FP}+S_{\mathrm{LCO}}+S_{ext}\;.  \label{sl}
\end{equation}
turns out thus to obey the following identities:

\begin{itemize}
\item  The Slavnov-Taylor identity 
\begin{equation}
\mathcal{S}(\Sigma )=0\;,  \label{st}
\end{equation}
\begin{equation}
\mathcal{S}(\Sigma )=\int d^{4}x\left( \frac{\delta \Sigma }{\delta A_{\mu
}^{a}}\frac{\delta \Sigma }{\delta \Omega ^{a\mu }}+\frac{\delta \Sigma }{%
\delta L^{a}}\frac{\delta \Sigma }{\delta c^{a}}+b^{a}\frac{\delta \Sigma }{%
\delta \overline{c}^{a}}+\partial _{\mu }J\frac{\delta \Sigma }{\delta \eta
_{\mu }}+\eta ^{\mu }\frac{\delta \Sigma }{\delta \tau ^{\mu }}-Jc^{a}\frac{%
\delta \Sigma }{\delta b^{a}}\right) \;  \label{sto}
\end{equation}

\item  The Landau gauge condition 
\begin{equation}
\frac{\delta \Sigma }{\delta b^{a}}=\partial _{\mu }A^{\mu a}\;,
\label{gcond}
\end{equation}
and the antighost Ward identity 
\begin{equation}
\frac{\delta \Sigma }{\delta \overline{c}^{a}}+\partial _{\mu }\frac{\delta
\Sigma }{\delta \Omega _{\mu }^{a}}=0\;,  \label{agh}
\end{equation}

\item  The ghost Ward identity \cite{bps,book} 
\begin{equation}
\mathcal{G}^{a}\Sigma =\Delta _{\mathrm{cl}}^{a}\;,  \label{lghi}
\end{equation}

where 
\begin{equation}
\mathcal{G}^{a}\Sigma =\int d^{4}x\left( \frac{\delta \Sigma }{\delta c^{a}}%
+gf^{abc}\overline{c}^{b}\frac{\delta \Sigma }{\delta b^{c}}-\tau _{\mu }%
\frac{\delta \Sigma }{\delta \Omega _{\mu }^{a}}\right) \;,  \label{gop}
\end{equation}
and 
\begin{equation}
\Delta _{\mathrm{cl}}^{a}=\int d^{4}x\left( gf^{abc}\Omega _{\mu
}^{b}A^{c\mu }-gf^{abc}L^{b}c^{c}+\eta ^{\mu }A_{\mu }^{a}\right) \;.
\label{dcl}
\end{equation}
Notice that expression $\left( \ref{dcl}\right) $, being purely linear in
the quantum fields, is a classical breaking. It is remarkable that the ghost
Ward identity can be established also in the presence of the external
sources $\left( J,\eta ^{\mu },\tau ^{\mu }\right) $. As we shall see, this
identity will play a fundamental role for the algebraic proof of the
relation $\left( \ref{ga2}\right) $.
\end{itemize}

\subsection{Algebraic characterization of the general local counterterm}

We are now ready to analyse the structure of the most general local
counterterm compatible with the identities $\left( \ref{st}\right) -$ $%
\left( \ref{lghi}\right) $. Let us begin by displaying the quantum numbers
of all fields and sources, namely 
\begin{equation}
\begin{tabular}{|l|l|l|l|l|l|l|l|l|l|}
\hline
& $A_{\mu }^{a}$ & $c^{a}$ & $\overline{c}^{a}$ & $b^{a}$ & $L^{a}$ & $%
\Omega _{\mu }^{a}$ & $J$ & $\eta ^{\mu }$ & $\tau ^{\mu }$ \\ \hline
Gh. number & $0$ & $1$ & $-1$ & $0$ & $-2$ & $-1$ & $0$ & $-1$ & $-2$ \\ 
\hline
Dimension & $1$ & $0$ & $2$ & $2$ & $4$ & $3$ & $2$ & $3$ & $3$ \\ \hline
\end{tabular}
\label{t1}
\end{equation}
In order to characterize the most general invariant counterterm which can be
freely added to all orders of perturbation theory, we perturb the classical
action $\Sigma $ by adding an arbitrary integrated local polynomial $\Sigma
^{\mathrm{count}}$ in the fields and external sources of dimension bounded
by four and with zero ghost number, and we require that the perturbed action 
$(\Sigma +\varepsilon \Sigma ^{\mathrm{count}})$ satisfies the same Ward
identities and constraints as $\Sigma $ to the first order in the
perturbation parameter $\varepsilon ,$ \textit{i.e.}

\begin{eqnarray}
\mathcal{S}(\Sigma +\varepsilon \Sigma ^{\mathrm{count}}) &=&0+O(\varepsilon
^{2})\;,  \nonumber \\
\frac{\delta (\Sigma +\varepsilon \Sigma ^{\mathrm{count}})}{\delta b^{a}}
&=&\partial ^{\mu }A_{\mu }^{a}+O(\varepsilon ^{2})\;,  \nonumber \\
\left( \frac{\delta }{\delta \overline{c}^{a}}+\partial _{\mu }\frac{\delta 
}{\delta \Omega _{\mu }^{a}}\right) (\Sigma +\varepsilon \Sigma ^{\mathrm{%
count}}) &=&0\;+O(\varepsilon ^{2})\;,  \nonumber \\
\mathcal{G}^{a}(\Sigma +\varepsilon \Sigma ^{\mathrm{count}}) &=&\Delta _{%
\mathrm{cl}}^{a}\;+O(\varepsilon ^{2})\;.  \label{cont-1}
\end{eqnarray}
This amounts to impose the following conditions on $\Sigma ^{\mathrm{count}}$

\begin{eqnarray}
\mathcal{B}_{\Sigma }\Sigma ^{\mathrm{count}} &=&0\;,  \label{st-c} \\
\mathcal{B}_{\Sigma } &=&\int d^{4}x\left( \frac{\delta \Sigma }{\delta
A_{\mu }^{a}}\frac{\delta }{\delta \Omega ^{a\mu }}+\frac{\delta \Sigma }{%
\delta \Omega ^{a\mu }}\frac{\delta }{\delta A_{\mu }^{a}}+\frac{\delta
\Sigma }{\delta L^{a}}\frac{\delta }{\delta c^{a}}+\frac{\delta \Sigma }{%
\delta c^{a}}\frac{\delta }{\delta L^{a}}\right. \;  \nonumber \\
&&\left. +b^{a}\frac{\delta }{\delta \overline{c}^{a}}+\partial _{\mu }J%
\frac{\delta }{\delta \eta _{\mu }}+\eta ^{\mu }\frac{\delta }{\delta \tau
^{\mu }}-Jc^{a}\frac{\delta }{\delta b^{a}}\right) \;,  \nonumber
\end{eqnarray}
and 
\begin{equation}
\frac{\delta \Sigma ^{\mathrm{count}}}{\delta b^{a}}=0\;,  \label{bc}
\end{equation}
\begin{equation}
\frac{\delta \Sigma ^{\mathrm{count}}}{\delta \overline{c}^{a}}+\partial
_{\mu }\frac{\delta \Sigma ^{\mathrm{count}}}{\delta \Omega _{\mu }^{a}}=0\;,
\label{a-c}
\end{equation}
\begin{equation}
\mathcal{G}^{a}\Sigma ^{\mathrm{count}}=0\;.  \label{gh-c}
\end{equation}
From equations $\left( \ref{bc}\right) $ and $\left( \ref{a-c}\right) $ it
follows that $\Sigma ^{\mathrm{count}}$ does not depend on the Lagrange
multiplier field $b^{a}$ and that the antighost $\overline{c}^{a}$ enters
only through the combination $\widehat{\Omega }_{\mu }^{a}=\left( \Omega
_{\mu }^{a}+\partial _{\mu }\overline{c}^{a}\right) $. As a consequence, $%
\Sigma ^{\mathrm{count}}$\ can be parametrized as follows 
\begin{eqnarray}
\Sigma ^{\mathrm{count}} &=&S^{\mathrm{count}}(A)  \nonumber \\
&&+\int d^{4}x\left( \frac{a_{1}}{2}f^{abc}L^{a}c^{b}c^{c}+a_{2}\widehat{%
\Omega }_{\mu }^{a}\partial ^{\mu }c^{a}+a_{3}f^{abc}\widehat{\Omega }_{\mu
}^{a}A_{\mu }^{b}c^{c}+\frac{a_{4}}{2}\xi J^{2}\right)  \nonumber \\
&&+\int d^{4}x\left( \frac{a_{5}}{2}JA_{\mu }^{a}A^{a\mu }+a_{6}\eta ^{\mu
}A_{\mu }^{a}c^{a}+\frac{a_{7}}{2}\tau ^{\mu }f^{abc}A_{\mu
}^{a}c^{b}c^{c}+a_{8}\tau ^{\mu }\partial _{\mu }c^{a}c^{a}\right) \;, 
\nonumber \\
&&  \label{ct1}
\end{eqnarray}
where $a_{i},\;i=1,..8,$ are free parameters and $S^{\mathrm{count}}(A)$
depends only on the gauge fields $A_{\mu }^{a}$. From the ghost Ward
identity condition $\left( \ref{gh-c}\right) $ it follows that

\begin{eqnarray}
a_{1} &=&a_{3}=a_{6}=a_{7}=0\;,  \label{gh-r} \\
a_{8} &=&-a_{2}\;.  \nonumber
\end{eqnarray}
The vanishing of the coefficient $a_{1}$ expresses the absence of the
counterterm $f^{abc}L^{a}c^{b}c^{c}$. Also, $a_{3}=0$ states the
nonrenormalization of the ghost-antighost-gluon vertex, stemming from the
transversality of the Landau propagator and from the factorization of the
ghost momentum. As shown in \cite{bps}, these features are related to a set
of nonrenormalization theorems of the Landau gauge. Furthermore, the
vanishing of $a_{6}$ implies the ultraviolet finiteness of the operator $%
A_{\mu }c$. It is easy to check indeed that at one loop order the 1PI
amputated Green function $\left\langle A_{\mu }(x)c(x)\;A_{\nu }(y)\overline{%
c}(z)\right\rangle _{1PI}$ with the insertion of $A_{\mu }c$ is not
divergent, due to the transversality of the Landau propagator. These
finiteness properties extend to all orders, due to the ghost Ward identity $%
\left( \ref{lghi}\right) $. It remains now to work out the condition $\left( 
\ref{st-c}\right) $. Making use of the well known results on the cohomology
of Yang-Mills theory \cite{book,bbh}, it turns out that the condition $%
\left( \ref{st-c}\right) $ implies that the coefficient $a_{5}$ is related
to $a_{2},$%
\begin{equation}
a_{5}=a_{2}\;,  \label{a5}
\end{equation}
and that $S^{\mathrm{count}}(A)$ can be written as

\begin{eqnarray}
S^{\mathrm{count}}(A) &=&\rho S_{YM}(A)+a_{2}\int d^{4}xA_{\mu }^{a}\frac{%
\delta S_{YM}(A)}{\delta A_{\mu }^{a}}\;,  \nonumber \\
S_{YM}(A) &=&-\frac{1}{4}\int d^{4}xF_{\mu \nu }^{a}F^{a\mu \nu }\;,
\label{sca}
\end{eqnarray}
where $\rho $ is a free parameter. In summary, the most general local
counterterm compatible with the Ward identities $\left( \ref{st}\right) -$ $%
\left( \ref{lghi}\right) $ contains three independent parameters $\rho $, $%
a_{2}$, $a_{4}$, and reads

\begin{eqnarray}
\Sigma ^{\mathrm{count}} &=&\rho S_{YM}(A)+a_{2}\int d^{4}xA_{\mu }^{a}\frac{%
\delta S_{YM}(A)}{\delta A_{\mu }^{a}}  \nonumber \\
&&+\int d^{4}x\left( a_{2}\left( \Omega _{\mu }^{a}+\partial _{\mu }%
\overline{c}^{a}\right) \partial ^{\mu }c^{a}+\frac{a_{4}}{2}\xi J^{2}+\frac{%
a_{2}}{2}JA_{\mu }^{a}A^{a\mu }-a_{2}\tau ^{\mu }\partial _{\mu
}c^{a}c^{a}\right) \;.  \nonumber \\
&&  \label{ctf}
\end{eqnarray}
It is apparent from the above expression that the parameters $\rho $ and $%
a_{2}$ are related to the renormalization of the gauge coupling $g$ and of
the gauge fields $A_{\mu }^{a}$, while the parameter $a_{4}$ corresponds to
the renormalization of $\xi $. It should be remarked also that the
coefficient of the counterterm $JA_{\mu }^{a}A^{a\mu }$ is given by $a_{2}$.
This means that the renormalization of the external source $J$, and thus of
the composite operator $A_{\mu }^{a}A^{a\mu }$, can be expressed as a
combination of the renormalization constants of the gauge coupling and of
the gauge fields. As we shall see in the next section, this property will
lead to the eq.$\left( \ref{ga2}\right) $.

\subsection{Stability and renormalization constants}

Having found the most general local counterterm compatible with all Ward
identities, it remains to discuss the stability \cite{book} of the classical
starting action, \textit{i.e.} to check that $\Sigma ^{\mathrm{count}}$ can
be reabsorbed in the starting action $\Sigma $ by means of a multiplicative
renormalization of the coupling constant $g$, the parameter $\xi $, the
fields $\left\{ \phi =A,c,\overline{c},b\right\} $ and the sources $\left\{
\Phi =J,\eta ,\tau ,L,\Omega \right\} $, namely

\begin{equation}
\Sigma (g,\xi ,\phi ,\Phi )+\varepsilon \Sigma ^{\mathrm{count}}=\Sigma
(g_{0},\xi _{0},\phi _{0},\Phi _{0})+O(\varepsilon ^{2})\;,  \label{stab}
\end{equation}
where, adopting the same conventions of \cite{gr}

\begin{eqnarray}
g_{0} &=&Z_{g}g\;,  \label{z} \\
\xi _{0} &=&Z_{\xi }\xi \;,  \nonumber \\
\phi _{0} &=&Z_{\phi }^{1/2}\phi \;,  \nonumber \\
\Phi _{0} &=&Z_{\Phi }\Phi \;.  \nonumber
\end{eqnarray}
As already mentioned, the parameters $\rho $ and $a_{2}$ are related to the
renormalization of $g$ and $A_{\mu }^{a}$, according to 
\begin{eqnarray}
Z_{g} &=&1-\varepsilon \frac{\rho }{2}\;,  \label{zgza} \\
Z_{A}^{1/2} &=&1+\varepsilon \left( a_{2}+\frac{\rho }{2}\right) \;. 
\nonumber
\end{eqnarray}
Concerning the other fields, it is almost immediate to verify that they are
renormalized as follows

\begin{equation}
Z_{b}=Z_{A}^{-1}\;,  \label{zb}
\end{equation}
and 
\begin{equation}
Z_{\overline{c}}=Z_{c}=Z_{g}^{-1}Z_{A}^{-1/2}\;.  \label{zc}
\end{equation}
Similar relations are easily found for the sources. In particular, for the
source $J$ and for the parameter $\xi $ one has

\begin{equation}
Z_{J}=Z_{A^{2}}=Z_{g}Z_{A}^{-1/2}\;,  \label{zj}
\end{equation}
and

\begin{equation}
Z_{\xi }=1+\varepsilon \left( a_{4}+2a_{2}+2\rho \right) =\left(
1+\varepsilon a_{4}\right) Z_{g}^{-2}Z_{A}\;.  \label{zx}
\end{equation}
We see therefore that the relation 
\begin{equation}
\gamma _{A^{2}}=-\left( \frac{\beta (a)}{a}+\gamma _{A}(a)\right) \;,
\label{frel}
\end{equation}
follows from eq.$\left( \ref{zj}\right) $. Concerning now the eq.$\left( \ref
{rel}\right) $ for the ghost anomalous dimension, it is a direct consequence
of eq.$\left( \ref{zc}\right) $.

Summarizing, we have been able to give a purely algebraic proof of the
relationship $\left( \ref{ga2}\right) $. This means that the anomalous
dimension $\gamma _{A^{2}}$ of the composite operator $A_{\mu }^{a}A^{a\mu
\;}$is not an independent parameter for Yang-Mills theory in the Landau
gauge.

\section{Acknowledgments}

D. Dudal would like to thank J. Gracey for sharing his results on the
possible existence of the relations $\left( \ref{ri}\right) $ and $\left( 
\ref{ri1}\right) $ prior to the publication of \cite{gr}. The Conselho
Nacional de Desenvolvimento Cient\'{i}fico e Tecnol\'{o}gico CNPq-Brazil,
the Funda{\c{c}}{\~{a}}o de Amparo {\ a Pesquisa do Estado do Rio de Janeiro
(Faperj) and the SR2-UERJ are acknowledged for the financial support. }

\end{document}